\documentclass[twocolumn,superscriptaddress,groupedaddress,amsmath,amssymb,aps,pra,floatfix]{revtex4-1}

\usepackage{comment}
\usepackage{graphicx}
\usepackage{dcolumn}
\usepackage{bm}

\usepackage[utf8]{inputenc}
\usepackage[T1]{fontenc}

\usepackage{braket}
\usepackage{siunitx}
\usepackage{xcolor}

\usepackage[normalem]{ulem}

\def\Rb87{^{87}\mathrm{Rb}}                             
\def\He4{^{4}\mathrm{He}}     

\begin{document}

\title{Interpretable Neural Network Quantum States for Solving the Steady States of the Nonlinear Schrödinger Equation}

\author{Mingshu Zhao}

\author{Zhanyuan Yan}
 \email{yanzhanyuan@ncepu.edu.cn}
\affiliation{Hebei Key Laboratory of Physics and Energy Technology, Department of Mathematics and Physics, North China Electric Power University, Baoding, Hebei 071003, China}

\date{\today} 

\begin{abstract}
The nonlinear Schrödinger equation (NLSE) underpins nonlinear wave phenomena in optics, Bose-Einstein condensates, and plasma physics, but computing its excited states remains challenging due to nonlinearity-induced non-orthonormality. Traditional methods like imaginary time evolution work for ground states but fail for excited states. We propose a neural network quantum state (NNQS) approach, parameterizing wavefunctions with neural networks to directly minimize the energy functional, enabling computation of both ground and excited states. By designing compact, interpretable network architectures, we obtain analytical approximation of solutions. We apply the solutions to a case of spatiotemporal chaos in the NLSE, demonstrating its capability to study complex chaotic dynamics. This work establishes NNQS as a tool for bridging machine learning and theoretical studies of chaotic wave systems.
\end{abstract}

\maketitle
\section{Introduction}

The nonlinear Schrödinger equation (NLSE) governs wave phenomena across diverse systems, including optical solitons~\cite{kartashov2011solitons}, Bose-Einstein condensates (BECs)~\cite{becker2008oscillations}, and plasmas~\cite{kuznetsov1986soliton}. In BECs, the Gross-Pitaevskii equation (GPE)~\cite{pitaevskii2016bose}—a canonical NLSE—defines the ground state as the condensate’s macroscopic wavefunction, while excited states encode collective modes such as dark solitons~\cite{becker2008oscillations} and vortex lattices~\cite{schweikhard2004vortex}. These stationary states underpin far-from-equilibrium dynamics. On one hand, mixing ground and excited states in harmonic traps triggers spatiotemporal chaos~\cite{zhao2025spatiotemporal}; on the other, vortex interactions drive turbulent cascades~\cite{zhao2025kolmogorov}.
Understanding such dynamics hinges on resolving excited states, yet their computation remains challenging due to the nonlinear system’s lack of orthonormal eigenmodes.

Methods like imaginary time evolution excel at finding ground states by damping high-energy modes but struggle with excited states unless restrictive symmetries (e.g., even potentials with odd initial states) are imposed~\cite{bao2006efficient}.
 In addition, some continuation algorithms~\cite{chang2007adaptive,chen2011exploring}  and Newton-based
iterative algorithm~\cite{marojevic2013energy} are also designed to compute excited states of NLSE. However, the convergence of these methods depend on the choice of initial data.
Recently, an optimization trajectory has been designed to stably search for the excited states of the NLSE~\cite{liu2023constrained}. While powerful, this technique raises a natural question: If solving for excited states is fundamentally an optimization problem, can neural networks (NNs) provide a more flexible and user-friendly framework? In this work, we address this question by parameterizing the wavefunction with NNs and optimizing it through direct energy functional minimization to solve the steady states.

Parametrizing wavefunctions with neural networks is not a new idea. The framework of neural network quantum states (NNQS) has become a powerful paradigm for quantum many-body systems, as demonstrated in studies of lattice models and spin systems~\cite{medvidovic2024neural,lange2024architectures}.  In linear Schrödinger equations, NNQS parameterizes wavefunctions to directly minimize energy functionals, enabling computation of ground states~\cite{medvidovic2024neural,lange2024architectures} and time-dependent dynamics~\cite{gutierrez2022real}. Recent work has extended NNQS to nonlinear systems like the GPE, successfully obtaining ground states~\cite{bao2025computing}. However, no prior study of NLSE has utilized NNQS to compute excited states, despite the broader success of neural networks in solving differential equations via direct functional optimization~\cite{poudel2024novel}. This gap limits the framework’s utility for studying nonlinear dynamics like chaos or turbulence, where excited states play a central role.  
 Here, we develop an NNQS framework to solve excited states of the NLSE and demonstrate its capability to resolve chaotic dynamics in a 1D harmonic trap.

Unlike mesh-based numerical methods (e.g., finite difference~\cite{morton2005numerical} or finite elements~\cite{johnson2009numerical}), NNQS inherently produce analytical, mesh-free solutions when constructed using elementary activation functions like $\tanh{(x)}$. 
While the outputs of finite-layer neural networks are technically analytical, their complexity often obscures physical interpretability. To address this, we reduce the number of hidden layers and nodes to distill the optimized NNQS into simplified analytical forms. We demonstrate this framework through a detailed case study of stationary states in a 1D harmonically trapped GPE. The resulting approximations not only enhance interpretability but also serve as practical bases for Galerkin approximations~\cite{galerkin1915sterzhni} in nonlinear dynamics~\cite{bland2018probing, podivilov2019hydrodynamic} or Bogoliubov–de Gennes (BdG) analyses of excitations~\cite{pitaevskii2016bose}. 

The remainder of this paper is structured as follows. We first introduce the NNQS framework and its numerical implementation for solving the steady states of the NLSE. Next, we compute the ground and excited states of a harmonically trapped 1D GPE and validate the results against established methods. We then employ a simplified NN to distill the NNQS solutions into interpretable forms. Leveraging these excited states, we investigate spatiotemporal chaos in the system by simulating mixed-state dynamics. Finally, we conclude with a discussion of broader implications and future applications of interpretable NNQS in nonlinear physics.

\section{Numerical Methods}

\subsection{Variational method}

The NLSE is given by:
\begin{equation}
i \frac{\partial \psi}{\partial t} = -\frac{1}{2} \nabla^2 \psi + V(\mathbf{r}) \psi + g |\psi|^2 \psi
\end{equation}
where $ \psi(\mathbf{r}, t) $ is the wavefunction, $ V(\mathbf{r}) $ is the potential, and $ g $ is the nonlinearity parameter. For steady state solutions, we look for time-independent solutions of the form $\psi(\mathbf{r}, t) = \psi(\mathbf{r}) e^{-i\mu t}$, where $\mu$ is the nonlinear eigenvalue. Substituting this form into the NLSE leads to the time-independent NLSE:
\begin{equation}
\mu \psi = -\frac{1}{2} \nabla^2 \psi + V(\mathbf{r}) \psi + g |\psi|^2 \psi
\end{equation}

The variational method leverages the equivalence between the steady state solutions of the NLSE and the saddle point solutions of the associated energy functional  $E[\psi, \psi^*]$ :
\begin{equation}
E[\psi, \psi^*] = \int \left( \frac{1}{2} |\nabla \psi|^2 + V(\mathbf{r}) |\psi|^2 + \frac{g}{2} |\psi|^4 \right) d\mathbf{r}
\end{equation}
with the normalization constraint $\int |\psi|^2 d\mathbf{r} = 1$.

To find the saddle point solutions, we consider the functional derivative of $ E[\psi, \psi^*]$ with respect to $\psi^*$ and introduce the Lagrange multiplier $\mu$ to enforce the normalization constraint. The Euler-Lagrange equation is then given by:
\begin{equation}
\frac{\delta}{\delta \psi^*} \left\{ E[\psi, \psi^*] - \mu \left(\int |\psi|^2 d\mathbf{r} -1 \right)\right\} = 0
\end{equation}
This leads to:
\begin{equation}
-\frac{1}{2} \nabla^2 \psi + V(\mathbf{r}) \psi + g |\psi|^2 \psi - \mu \psi = 0
\end{equation}
This equation confirms that the steady state solutions of the NLSE correspond to the saddle points of the energy functional.
The ground state, $\psi_0$, is the energy-minimizing saddle point of the functional, representing the system's most stable configuration. Other saddle points correspond to excited states, which we order hierarchically by energy: $\psi_1$ (first excited state), $\psi_2$ (second excited state), and so on.

Since we are interested in the steady state solutions, the wavefunction can be considered real. This simplifies the energy functional and the corresponding Euler-Lagrange equation. For a real wavefunction  $\psi$, the energy functional reduces to:
\begin{equation}
E[\psi] = \int \left( \frac{1}{2} (\nabla \psi)^2 + V(\mathbf{r}) \psi^2 + \frac{g}{2} \psi^4 \right) d\mathbf{r}
\label{eq:Energy_functional}
\end{equation}
By minimizing this functional, we obtain the ground state, and by incorporating additional constraints, we can target excited states as well.

\subsection{Neural Network Representation and Optimization}

\begin{figure}[tb]
\centering
\includegraphics{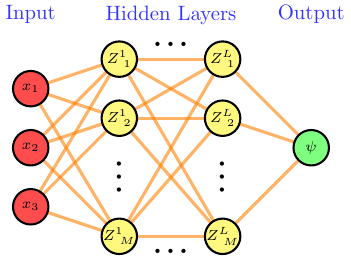}
\caption{Schematic of the NNQS framework. The MLP maps spatial coordinates $\bm{r} = (x_1, x_2, x_3)$ (red) through hidden layers (yellow) to output the wavefunction $\psi(\bm{r})$ (green). The network is trained by minimizing a composite loss functional (Eq.~\eqref{eq:total_loss}) to compute steady-state solutions of the NLSE.}
\label{fig:nn_architecture}
\end{figure}

We propose a NNQS framework to compute steady-state solutions of the NLSE. The wavefunction $\psi(\mathbf{r})$ is parameterized by a multilayer perceptron (MLP)~\cite{rumelhart1986learning} that maps spatial coordinates $\mathbf{r}$ to real-valued outputs. The network architecture consists of an input layer for $\mathbf{r}$, hidden layers with nonlinear activation functions, and an output layer yielding $\psi(\mathbf{r})$, as shown schematically in Fig.~\ref{fig:nn_architecture}.

The network is optimized by minimizing a composite loss functional $\mathcal{L}$ that combines the NLSE energy functional with physical constraints. The total loss is defined as:
\begin{equation}
\label{eq:total_loss}
\mathcal{L} = E[\psi] + \lambda_{\text{norm}} \mathcal{L}_{\text{norm}} + \lambda_{\text{excited}} \mathcal{L}_{\text{excited}} + \lambda_{\text{NLSE}} \mathcal{L}_{\text{NLSE}},
\end{equation}
where $E[\psi]$ is the NLSE energy functional from Eq.~\eqref{eq:Energy_functional}. The normalization penalty $\mathcal{L}_{\text{norm}} = \left( \int \psi^2 d\mathbf{r} - 1 \right)^2$ enforces normalization of the wavefunction. 

For excited states, $\mathcal{L}_{\text{excited}}$ penalizes overlaps with all prior lower-energy states $\{\psi_i\}_{i=0}^{n_{\text{prior}}}$ when they exceed a threshold $\epsilon$. For each state $\psi_i$, the overlap is computed as $\mathcal{O}_i = \left|\int \psi_i(\mathbf{r}) \psi(\mathbf{r}) \, d\mathbf{r}\right|$, and the penalty is applied conditionally:
\[
\mathcal{L}_{\text{excited}} = \sum_{i=0}^{n_{\text{prior}}} 
\begin{cases} 
\mathcal{O}_i, & \text{if } \mathcal{O}_i \geq \epsilon, \\
0, & \text{otherwise},
\end{cases}
\]  
where $\epsilon$ is the nonlinearity-dependent threshold. This ensures the current state $\psi$ is distinct from all previously found lower-energy prior states while respecting the non-orthonormal nature of nonlinear eigenstates.

The NLSE residual term $$\mathcal{L}_{\text{NLSE}} = \int \left|\mathcal{N}\left(-\frac{1}{2} \nabla^2 \psi + V(\mathbf{r}) \psi + g \psi^3 \right) - \psi\right| d\mathbf{r}$$ ensures consistency with the steady-state NLSE, where $\mathcal{N}$ denotes wavefunction normalization. While not strictly necessary, including the NLSE residual loss accelerates convergence during optimization.
Weighting coefficients $\lambda_{\text{norm}}$, $\lambda_{\text{excited}}$, and $\lambda_{\text{NLSE}}$ in Eq.~\eqref{eq:total_loss} balance these constraints.

Network parameters $\theta$ are optimized using gradient-based methods such as stochastic gradient descent~\cite{bottou2010large} or Adam~\cite{kingma2014adam}. The workflow iteratively samples spatial coordinates $\mathbf{r}$, evaluates $\psi(\mathbf{r})$ by forward passing the NN and its gradients via automatic differentiation~\cite{baydin2018automatic} or finite difference method, and updates $\theta$ to minimize $\mathcal{L}$. Convergence to a steady state occurs when $\mathcal{L}$ stabilizes, with the global minimum corresponding to the stationary states.

It is worth noting that the optimization of the ground state is generally slower than imaginary time evolution, which can be interpreted as a form of natural gradient descent for quantum state optimization~\cite{stokes2020quantum}. However, implementing exact second-order methods, such as those based on the Fisher information matrix~\cite{amari2016information}, becomes computationally inefficient for neural network parameterizations. In practice, efficient approximations like the Kronecker-factored approximate curvature (K-FAC) method~\cite{martens2015optimizing} serve as viable alternatives. Nevertheless, for simplicity, we primarily use the Adam optimizer in this work.


\section{Stationary Solutions}

We investigate the one-dimensional NLSE in a harmonic potential $V(x) = \frac{1}{2}x^2$. Since this system models spatiotemporal chaos in atomic BECs, we treat the NLSE as the GPE with an interaction strength $g$. For typical experimental conditions~\cite{zhao2025spatiotemporal}, $g \leq 100$, and we present results for $g = 1$, $10$, and $100$.

The numerical simulations are implemented using PyTorch~\cite{paszke2017automatic} on a CPU. In our case, automatic differentiation does not outperform finite differences in computational efficiency, so we compute $\partial_x \psi$ using a second-order finite-difference scheme on a uniform grid of $4097$ points over the domain $x \in [-16, 16]$.

We employ a MLP with the following architecture. The input dimension is $1$ due to the one-dimensional nature of the problem. The network consists of seven layers with GeLU~\cite{hendrycks2016gaussian} activation functions: a linear layer mapping $1 \to 256$, followed by two linear layers $256 \to 256$, then a reduction $256 \to 128$, another layer $128 \to 128$, a further reduction $128 \to 64$, and finally a linear output layer $64 \to 1$. This structure balances expressive power and computational efficiency for our study.

\subsection{Ground State}
The ground state solutions are obtained by minimizing a loss function comprising solely the energy functional $E[\psi]$ and a normalization constraint term ($\mathcal L_{\text{norm}}$), excluding contributions from excited states ($\mathcal L_{\text{excited}}$) and NLSE residue ($\mathcal L_{\text{NLSE}}$). The normalization penalty strength is set to $\lambda_{\text{norm}}=100$ to strictly enforce particle number conservation. Optimization is performed using the Adam algorithm with a learning rate of $5 \times 10^{-5}$. 

\begin{figure}[tb]
\centering
\includegraphics{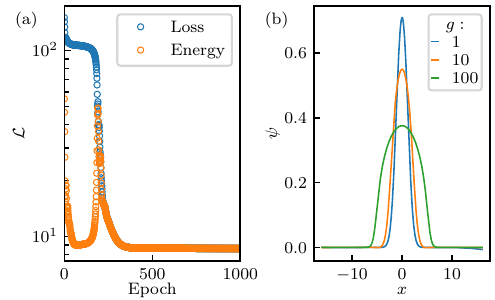}
\caption{Optimization history and ground state solutions. (a) Convergence history showing the loss (blue) and energy (orange) for $g=100$. (b) Optimized ground state profiles for various interaction strengths $g=1,10,100$.}
\label{fig:ground_state_results}
\end{figure}

Figure~\ref{fig:ground_state_results} presents the numerical results: panel (a) displays the evolution of both the loss function and energy as functions of training epochs for the case of $g=100$, while panel (b) shows the converged ground state solutions across different interaction strengths $g$. The obtained solutions agree with results from imaginary time evolution, validating our optimization approach.

During the initial optimization phase, the loss function exceeds the system energy due to non-normalized neural network outputs. This discrepancy resolves after approximately 200 epochs as the network converges to properly normalized states. Full convergence to ground state solutions is achieved within 500 epochs.


\subsection{First Excited State}

The calculation of first excited states builds upon the obtained ground state solutions. Our optimization scheme employs penalty strengths $\lambda_{\text{excited}}=20$ for excited state separation, $\lambda_{\text{norm}}=100$ for normalization, and $\lambda_{\text{NLSE}}=10$ for the governing equation constraint, with an overlap threshold $\epsilon=0.5$ in the excited state loss function. We maintain the Adam optimizer configuration with learning rate $5\times10^{-5}$ for consistency.

\begin{figure}[tb]
\centering
\includegraphics{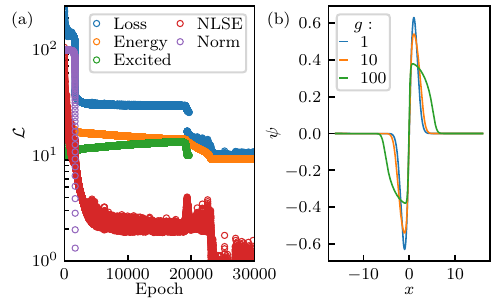}
\caption{Optimization history and first excited state solutions. (a) Convergence history showing the total loss (blue), energy (orange), excited state loss (green), NLSE residue loss (red) and normalization loss (purple) for $g=100$. (b) Optimized first excited state profiles for various interaction strengths $g=1,10,100$.}
\label{fig:excited1st_state_results}
\end{figure}

Figure~\ref{fig:excited1st_state_results}(a) reveals the optimization dynamics for $g=100$ through three characteristic phases. The initial normalization phase ($\sim$2000 epochs) shows the normalization error (purple markers) decaying to zero. Subsequently, the system enters an excited and ground state separation phase ($\sim$20000 epochs) where solutions remain in configuration space regions with significant ground state overlap. The final excited state convergence phase ($\sim$3000 epochs) begins with a sharp transition to excited-state-dominant configurations, evidenced by the excited state loss (green curve) abruptly dropping to zero, followed by energy decay (orange curve) to the first excited state energy level.
We observe that while the energy remains conserved after $25000$ epochs, the NLSE residual loss (red markers) oscillates persistently about unity. This behavior indicates that the current learning rate, though sufficient for first excited state searching, is too large for complete residual minimization. For improved accuracy in the solutions, subsequent optimization with a reduced learning rate would be necessary.

This multistage convergence profile, requiring approximately two orders of magnitude more epochs than ground state optimization, indicates the increased computational complexity of excited state calculations. The prolonged intermediate phase particularly reflects the challenging energy landscape navigation between ground and excited state configurations.

Figure~\ref{fig:excited1st_state_results}(b) presents the first excited state solutions, which match well with imaginary time evolution results using odd-parity initial states.

\subsection{Higher Excited States}

Higher excited states are inaccessible through standard imaginary time evolution, and their optimization presents significantly greater computational complexity. To mitigate this challenge, we employ an initialization using known solutions from other cases: For weakly nonlinear cases ($g=1$), we initialize the neural network with noninteracting excited state solutions, enabling rapid convergence. This solution then serves as the initial condition for progressively stronger interactions ($g=10$), and the latter one subsequently for the strongly interacting case ($g=100$), establishing an efficient hierarchical optimization strategy.

During optimization, we impose additional symmetry constraints on the excited states by introducing a penalty term proportional to $\mathcal{L}_{\text{sym}} = \text{mean}\left[(|\psi(x)| - |\psi(-x)|)^2\right]$. This symmetry-enforcing loss function serves two key purposes: (1) it accelerates convergence by guiding the optimization toward physically relevant solutions, and (2) it systematically eliminates unphysical asymmetric density $|\psi(x)|^2$ configurations from the solution space.

\begin{figure*}[t]
\centering
\includegraphics{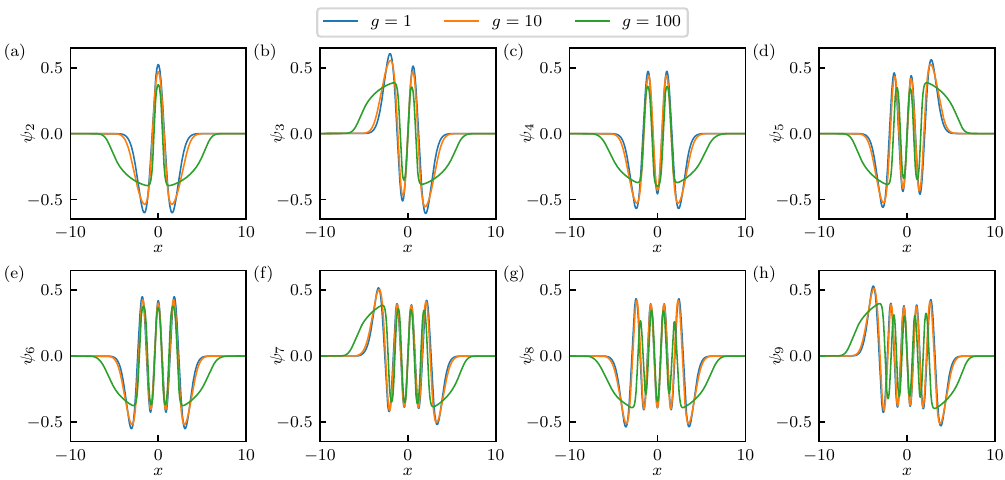}
\caption{Excited state solutions $\psi_i(x)$ from $i=2$ to $i=9$ for interaction strengths $g=1$ (blue), $10$ (orange), and $100$ (green).}
\label{fig:higher_excited_state_results}
\end{figure*}

Figure~\ref{fig:higher_excited_state_results} displays the excited states ($i=2,...,9$) for $g=1,10,100$, showing agreement with Ref.~\cite{liu2023constrained}. The solutions reveal two key features: (i) the number of nodes matches the noninteracting case, implying topological constraints persist despite nonlinear interactions, and (ii) strong interactions compress nodes toward the trap center (notably for $i=8,9$) while expanding the Thomas-Fermi envelope~\cite{pitaevskii2016bose}. The wavefunctions deviate significantly from Hermite-Gaussian forms~\cite{zwiebach2022mastering}, particularly at large $g$.

The nodal compression observed at large $g$ emerges from an energy trade-off: while compressed nodes increase kinetic energy through steeper gradients ($E_{\text{kin}}\sim\int|\nabla\psi|^2 dx$), this is outweighed by interaction energy reduction ($E_{\text{int}}\sim g\int|\psi|^4 dx$) as density redistributes outward.

While the NNQS solutions provide qualitatively understandable results, deriving approximate analytical expressions could yield deeper physical insights. We address this through simplified NN in the following section.

\section{Interprebility}


\begin{figure}[tbh]
\centering
\includegraphics{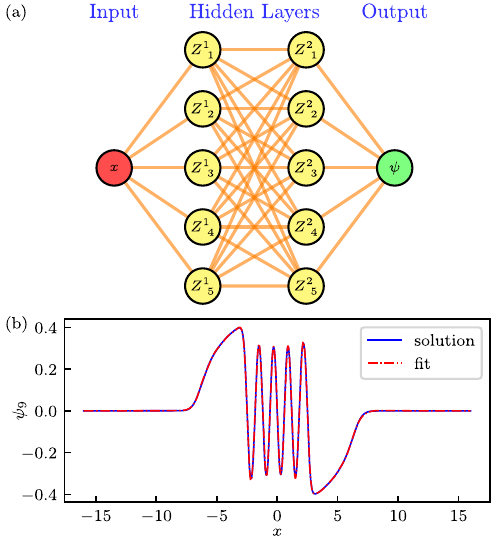}
\caption{Interpretable NNQS architecture and performance. 
    (a) Architecture of the compact NNQS model: Spatial coordinates $x$ (input, red) are mapped through two hidden layers (yellow, 5 nodes each, $\tanh$ activation) to output the wavefunction $\psi(x)$ (green). 
    (b) Performance of the compact NNQS model from (a): The fitted $\psi_9(x)$ (red dashed) for $g = 100$ closely matches the exact numerical solution [blue, from Fig.~\ref{fig:higher_excited_state_results}(h)].}
\label{fig:Interpret}
\end{figure}

Unlike conventional black-box neural networks, our NNQS method provides inherently interpretable solutions that surpass traditional grid-based algorithms. Grid methods are limited to discrete numerical results, whereas NNQS generates continuous wavefunctions across the entire spatial domain. However, the model producing the solution in Fig.~\ref{fig:higher_excited_state_results} contains approximately $2\times10^5$ parameters, making direct interpretation challenging. Specialized techniques are needed to extract meaningful physical insights from these complex solutions.

Several approaches have been developed to improve neural network interpretability in scientific applications. Notable examples include symbolic regression via genetic evolution to identify optimal analytical solutions~\cite{cranmer2023interpretable,makke2024interpretable,cranmer2020discovering}, and the recent Kolmogorov-Arnold network (KAN)~\cite{liu2024kan,liu2024kan2,kolmogorov1961representation} that employs B-spline functions~\cite{de1978practical} to represent solutions with at most two activation layers, enabling subsequent symbolic expression extraction. While these methods succeed when solutions can be expressed through simple elementary functions, they face limitations: symbolic regression often fails to find satisfactory approximations for complex cases, and while KANs reduce parameter counts, their B-spline-based results remain fundamentally difficult to interpret.

Motivated by theoretical studies of spinor interfaces in NLSE~\cite{takahashi2015nambu, indekeu2015static} - where ground state solutions frequently feature $\tanh$ profiles - we implement a compact neural network architecture with $\tanh$ activation [Fig.~\ref{fig:Interpret}(a)]. Strikingly, this minimal network, comprising just two hidden layers with five nodes each, accurately reproduces even our most challenging excited state solution [$\psi_9(x)$ at $g=100$, Fig.~\ref{fig:Interpret}(b)].

This minimal architecture contains only 46 parameters while maintaining a mean squared error of $6.1\times10^{-6}$ when fitting the solutions. Although not the absolute simplest possible configuration, this $\tanh$-activated network provides an effective balance between interpretability and accuracy. Further simplification could be achieved through alternative activation functions or pruning less influential nodes.

The small parameter count facilitates direct interpretation and serves as an ideal foundation for analytical methods like the Galerkin approximations~\cite{galerkin1915sterzhni,bland2018probing,podivilov2019hydrodynamic} or BdG analysis of stationary solutions.

\section{Spatiotemporal chaos}

Recent work has demonstrated spatiotemporal chaos emergence in a 1D harmonic trap when initializing the system with a superposition of ground and first excited states~\cite{zhao2025spatiotemporal}. In this regime, nonlinear interactions generate higher excited state modes through a direct-cascade-like mechanism~\cite{kolmogorov1995turbulence}. Notably, extended self-similarity (ESS)~\cite{benzi1993extended}—a hallmark of turbulent systems—has been observed in these dynamics. 

The present study investigates two key questions: (1) whether enhanced chaotic behavior emerges when initializing the system with superpositions of the ground state and higher odd-order excited states, and (2) whether the ESS phenomenon persists in these modified initial conditions. This will provide deeper understanding of how initial states affect wave chaos generation and the formation of ESS.

\subsection{Lyapunov Exponents}
The Lyapunov exponent~\cite{strogatz2018nonlinear} plays a crucial role in characterizing chaos, serving not only as an indicator of chaotic dynamics through its positivity but also as a measure of the GPE's validity for studying BEC dynamics~\cite{zhao2025spatiotemporal}. Fundamentally, the many-body Schrödinger equation is linear and thus cannot exhibit chaos; the observed chaotic behavior stems from the nonlinear interactions in the mean-field approximation. Consequently, the GPE is not expected to remain valid indefinitely once the system becomes chaotic~\cite{bvrezinova2012wave}. However, the atom number $N$ extends the GPE's validity beyond the typical Lyapunov time up to the Ehrenfest time $\tau_E = \ln(N)/\lambda$, where $\lambda$ is the Lyapunov exponent of the mean-field dynamics~\cite{ehrenfest1927bemerkung,berman1978condition,han2016ehrenfest,rammensee2018many,wanzenbock2021chaos}. 
Accordingly, our analysis proceeds in two stages: first computing Lyapunov exponents across parameter regimes, then using $\tau_E$ to bound our dynamical simulations' temporal domain for further structure function analysis~\cite{kolmogorov1995turbulence,zhao2025spatiotemporal}.

\begin{table*}[tb]
\vspace{0.5cm}
\caption{\label{tab:Lyapunov}%
Lyapunov exponents $\lambda$ for different 
 interaction strengths $g$ with initial states prepared as a superposition of the ground state and excited states $\psi_{2j+1}$, mixed according to ratio $\alpha$.}
\vspace{0.5cm}
\begin{ruledtabular}
\begin{tabular}{c c c c c c}
\textrm{} & 
\multicolumn{5}{c}{$g = 1,\ \alpha=0.1$} \\
\cline{1-6}
$j$ & 
\textrm{0} & 
\textrm{1} & 
\textrm{2} & 
\textrm{3} & 
\textrm{4} \\
\hline
$\lambda$ & $1(3) \times 10^{-6}$ & $1(8) \times 10^{-3}$  & $7(2) \times 10^{-3}$ 
 & $8(2) \times 10^{-3}$  & $1.0(2) \times 10^{-2}$\\
\hline
\textrm{} & 
\multicolumn{5}{c}{$g = 10,\ \alpha=0.1$} \\
\cline{1-6}
$j$ & 
\textrm{0} & 
\textrm{1} & 
\textrm{2} & 
\textrm{3} & 
\textrm{4} \\
\hline
$\lambda$ & $4.8(9) \times 10^{-3}$ & $1.1(7) \times 10^{-2}$  & $2(1) \times 10^{-2}$ 
 &$3(1) \times 10^{-2}$   &$4(1) \times 10^{-2}$\\
\hline
\textrm{} & 
\multicolumn{5}{c}{$g = 100,\ \alpha=0.1$} \\
\cline{1-6}
$j$ & 
\textrm{0} & 
\textrm{1} & 
\textrm{2} & 
\textrm{3} & 
\textrm{4} \\
\hline
$\lambda$ & $2.6(2) \times 10^{-2}$ & $6.0(8) \times 10^{-2}$ &  $7.8(7) \times 10^{-2}$
 & $9.7(8) \times 10^{-2}$  & $1.0(3) \times 10^{-1}$\\
\hline
\textrm{} & 
\multicolumn{5}{c}{$g = 1,\ \alpha=1$} \\
\cline{1-6}
$j$ & 
\textrm{0} & 
\textrm{1} & 
\textrm{2} & 
\textrm{3} & 
\textrm{4} \\
\hline
$\lambda$ & $2.2(2) \times 10^{-2}$ & $5.4(9) \times 10^{-2}$ & $6(1) \times 10^{-2}$  
 &  $2.8(3) \times 10^{-2}$ &$2.5(3) \times 10^{-2}$\\
\hline
\textrm{} & 
\multicolumn{5}{c}{$g = 10,\ \alpha=1$} \\
\cline{1-6}
$j$ & 
\textrm{0} & 
\textrm{1} & 
\textrm{2} & 
\textrm{3} & 
\textrm{4} \\
\hline
$\lambda$ & $1.1(1) \times 10^{-1}$ & $1.5(2) \times 10^{-1}$ &  $1.6(3) \times 10^{-1}$
 & $1.6(2) \times 10^{-1}$  & $1.4(3) \times 10^{-1}$\\
\hline
\textrm{} & 
\multicolumn{5}{c}{$g = 100,\ \alpha=1$} \\
\cline{1-6}
$j$ & 
\textrm{0} & 
\textrm{1} & 
\textrm{2} & 
\textrm{3} & 
\textrm{4} \\
\hline
$\lambda$ & $2.2(4) \times 10^{-1}$ &$3.2(5) \times 10^{-1}$   & $3.1(6) \times 10^{-1}$ 
 & $2.9(7) \times 10^{-1}$  &$2.5(6) \times 10^{-1}$\\
\end{tabular}
\end{ruledtabular}
\end{table*}

The initial wavefunction for real-time evolution is constructed as a superposition of states:
\begin{equation} 
\psi_{i}(x) = \frac{\psi_0(x) + \alpha \psi_{2j+1}(x)}{\sqrt{1+\alpha^2}}, 
\end{equation} 
where $\alpha$ controls the mixing ratio and $j=0,1,2,3,4$ indexes the odd-order excited states. The orthogonality between symmetric $\psi_0$ and antisymmetric $\psi_{2j+1}$ states guarantees the normalization factor takes this simple form.

We examine two distinct mixing regimes:
(i) Weak mixing ($\alpha=0.1$): The small excited-state admixture produces a smooth profile without phase discontinuities. This configuration can be experimentally prepared by engineering an external potential that molds the ground state into the desired profile, followed by a sudden quench to the pure harmonic trap potential.
(ii) Balanced mixing ($\alpha=1$): The equal superposition creates a $\pi$-phase jump at nodes, presenting greater experimental challenges due to the required precise phase control.

Table~\ref{tab:Lyapunov} summarizes the computed Lyapunov exponents $\lambda$ for various interaction strengths $g$, mixing parameters $\alpha$, and excited state combinations. The universally positive $\lambda$ values confirm the chaotic dynamics across all parameter regimes. 
Notably, strong mixing ($\alpha=1$) yields systematically larger Lyapunov exponents ($\lambda$), indicating enhanced chaotic dynamics. For weak mixing ($\alpha=0.1$), $\lambda$ increases with the order of the excited states in the initial superposition, as these states deviate more markedly from the ground-state configuration—a trend consistent with expectations.
However, in the strong mixing regime, the maximal $\lambda$ occurs for initial mixing with $\psi_5$, suggesting a nontrivial competition between inverse and direct energy cascades. We defer a detailed analysis of this complex dynamical interplay to future work.

\subsection{Extended self-similarity}

\begin{figure}[tbh]
\centering
\includegraphics{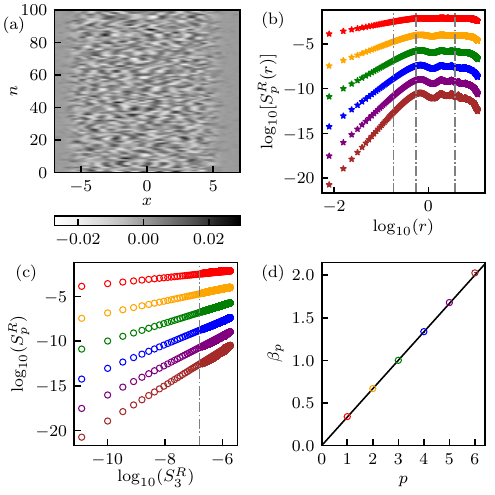}
\caption{Statistics of the initial $\psi_9$ mixing with $g=100,\ \alpha=0.1$.
(a) Fluctuating density sequence $\rho_n'(x)$. 
(b) Log-log plot of spatial density structure function $S_p^R(r)$ versus $r$ for $p=1 \dots 6$, with color coding as shown on the x-axis in panel (d). 
Grey dashed lines indicate positions of $r_0$, $r_1 = 3r_0$, and $r_2 = 20r_0$, respectively, where $r_0 = 1/\sqrt{2\mu_0}$ represents the ground-state healing length.
(c) Log-log plot of $S_p^R$ versus $S_3^R$. The grey dashed line indicates $S_3^R(r_0)$.
(d) ESS scaling exponents $\beta_p$, obtained from linear fits to the data in panel (c). The black line represents the K41 scaling law, $p/3$.}
\label{fig:ESS}
\end{figure}

We analyze ESS in the harmonic trapped BEC system through statistical properties of density fluctuations~\cite{zhao2025spatiotemporal}. The Kohn theorem~\cite{kohn1961cyclotron,brey1989optical,dobson1994harmonic} guarantees the separation of center-of-mass and relative motions in the harmonic trap. We leverage this property to assemble density profiles $\{\rho_n(x)\}$ with identical center-of-mass positions across different time snapshots, creating a proper ensemble for structure function analysis.

Following Reynolds decomposition principles~\cite{reynolds1895iv}, we separate the density field into mean and fluctuating components:
\begin{equation}
    \rho_n(x) = \langle \rho(x) \rangle + \rho'_n(x),
\end{equation}
where $\langle \rho(x) \rangle$ is the ensemble-averaged density and $\rho'_n(x)$ represents fluctuations at timestep $n$. Figure~\ref{fig:ESS}(a) displays fluctuation profiles for the strongly interacting regime ($g=100$) with weak mixing ($\alpha=0.1$) and initial state preparation involving $\psi_9$. Based on the Lyapunov exponent $\lambda=0.1$ from Table~\ref{tab:Lyapunov} and using atom numbers $N=10^5$, we estimate the GPE valid time as $\tau_E \approx 115$. This estimate constrains our statistical ensemble to $n \leq 100$ snapshots.

The spatial density increment $\delta \rho(r) = \rho'_n(x+r) - \rho'_n(x)$ serves as the fundamental quantity for statistical analysis. We compute the probability density functions $P[\delta \rho(r)]$ via histogramming and obtain the $p$-th order structure functions:
    \begin{equation}
        S_p^R(r) = \int \left|\delta \rho(r)\right|^p P[\delta \rho(r)] d[\delta \rho(r)]
        \label{Eq:struc_func}
    \end{equation}

Figure.~\ref{fig:ESS}(b) shows the $S_p^R(r)$. When direct $r$-scaling proves ambiguous, ESS reveals hidden scaling relations by comparing structure function of different orders:
\begin{equation}
    S_p^R(r) \propto S_3^R(r)^{\beta_p},
\end{equation}
where $\beta_p$ exponents characterize the inter-order scaling. As demonstrated in Fig.~\ref{fig:ESS}(c), this method effectively extends the scaling range to the entire dataset, even in cases with underdeveloped inertial ranges~\cite{benzi1993extended}. The resulting ESS scaling exponents $\beta_p$, shown in Fig.~\ref{fig:ESS}(d), exhibit excellent agreement with the K41 prediction $\beta_p = p/3$ (black line)~\cite{kolmogorov1941degeneration,kolmogorov1991local,kolmogorov1991dissipation}.

Our analysis reveals strong ESS across all examined cases, suggesting that ESS represents a universal property of the system's spatiotemporal chaos.

\section{Conclusion and Outlooks}

In this work, we develop a NNQS framework to compute stationary solutions of the NLSE through direct minimization of a composite energy functional. Our implementation employs a 7-layer MLP for the GPE with a 1D harmonic trap potential.  

To enhance interpretability, we reduce the NNQS complexity by fitting the solutions to a compact architecture with only 46 parameters, consisting of two layers (5 nodes per layer) with linear and $\tanh$ activation functions. This simplified model enables analytical approximation and provides physical insights into the solutions.  

Our investigation of spatiotemporal chaos employs superpositions of odd-order excited states with the ground state. The resulting chaotic dynamics are verified through two complementary signatures: (1) consistently positive Lyapunov exponents and (2) robust ESS across all examined cases.

The maximal Lyapunov exponent for equal mixing ($\alpha=1$) occurs unexpectedly when the excited state is $\psi_5$, suggesting nontrivial dynamics that merit further investigation. This raises a fundamental question: Under what mixing conditions does the system exhibit maximal chaos? 

For chaotic systems, several compelling directions emerge. 
First, computation of higher excited states could construct a complete basis for dynamic mode decomposition (DMD)~\cite{schmid2010dynamic,schmid2022dynamic}, analogous to techniques used in fluid turbulence analysis. Second, investigation of quasi-integrability in harmonically trapped GPE systems~\cite{bland2018probing,thomas2021thermalization} through systematic studies of initial state classification presents rich opportunities. Third, extension to higher-dimensional systems with controlled dissipation could probe turbulence formation.  

The quasi-integrability study warrants particular emphasis: by generating datasets from varied initial conditions (constrained to identical center-of-mass positions via Kohn's theorem), for a given chaotic density profile, we can develop classifiers to predict its corresponding initial state. This method quantifies the degree of integrability in a near-integrable system.

In the NNQS framework, simulating time-dependent dynamics for NLSE systems remains a key computational challenge. However, the inherent normalization-preserving structure of the approach indicates that diffusion-based algorithms could be particularly well-suited for ensuring numerical stability during evolution~\cite{orlova2023deep}.

The interpretable NNQS framework provides a powerful tool for investigating interface profiles in two-component BECs under external magnetic gradients. In such systems, exact analytical ground states are inaccessible, but NNQS enables numerical exploration while leveraging approximate analytical solutions to study interface modes—particularly in Rayleigh-Taylor unstable configurations~\cite{geng2024rayleigh}.

\begin{acknowledgments}
The authors thank Ian Spielman for helpful discussion. 
This work was supported by the Science and Technology Projects of China Southern Power Grid (YNKJXM20220050).
\end{acknowledgments}



\begin{thebibliography}{62}%
\makeatletter
\providecommand \@ifxundefined [1]{%
 \@ifx{#1\undefined}
}%
\providecommand \@ifnum [1]{%
 \ifnum #1\expandafter \@firstoftwo
 \else \expandafter \@secondoftwo
 \fi
}%
\providecommand \@ifx [1]{%
 \ifx #1\expandafter \@firstoftwo
 \else \expandafter \@secondoftwo
 \fi
}%
\providecommand \natexlab [1]{#1}%
\providecommand \enquote  [1]{``#1''}%
\providecommand \bibnamefont  [1]{#1}%
\providecommand \bibfnamefont [1]{#1}%
\providecommand \citenamefont [1]{#1}%
\providecommand \href@noop [0]{\@secondoftwo}%
\providecommand \href [0]{\begingroup \@sanitize@url \@href}%
\providecommand \@href[1]{\@@startlink{#1}\@@href}%
\providecommand \@@href[1]{\endgroup#1\@@endlink}%
\providecommand \@sanitize@url [0]{\catcode `\\12\catcode `\$12\catcode `\&12\catcode `\#12\catcode `\^12\catcode `\_12\catcode `\%12\relax}%
\providecommand \@@startlink[1]{}%
\providecommand \@@endlink[0]{}%
\providecommand \url  [0]{\begingroup\@sanitize@url \@url }%
\providecommand \@url [1]{\endgroup\@href {#1}{\urlprefix }}%
\providecommand \urlprefix  [0]{URL }%
\providecommand \Eprint [0]{\href }%
\providecommand \doibase [0]{https://doi.org/}%
\providecommand \selectlanguage [0]{\@gobble}%
\providecommand \bibinfo  [0]{\@secondoftwo}%
\providecommand \bibfield  [0]{\@secondoftwo}%
\providecommand \translation [1]{[#1]}%
\providecommand \BibitemOpen [0]{}%
\providecommand \bibitemStop [0]{}%
\providecommand \bibitemNoStop [0]{.\EOS\space}%
\providecommand \EOS [0]{\spacefactor3000\relax}%
\providecommand \BibitemShut  [1]{\csname bibitem#1\endcsname}%
\let\auto@bib@innerbib\@empty
\bibitem [{\citenamefont {Kartashov}\ \emph {et~al.}(2011)\citenamefont {Kartashov}, \citenamefont {Malomed},\ and\ \citenamefont {Torner}}]{kartashov2011solitons}%
  \BibitemOpen
  \bibfield  {author} {\bibinfo {author} {\bibfnamefont {Y.~V.}\ \bibnamefont {Kartashov}}, \bibinfo {author} {\bibfnamefont {B.~A.}\ \bibnamefont {Malomed}},\ and\ \bibinfo {author} {\bibfnamefont {L.}~\bibnamefont {Torner}},\ }\href@noop {} {\bibfield  {journal} {\bibinfo  {journal} {Reviews of Modern Physics}\ }\textbf {\bibinfo {volume} {83}},\ \bibinfo {pages} {247} (\bibinfo {year} {2011})}\BibitemShut {NoStop}%
\bibitem [{\citenamefont {Becker}\ \emph {et~al.}(2008)\citenamefont {Becker}, \citenamefont {Stellmer}, \citenamefont {Soltan-Panahi}, \citenamefont {D{\"o}rscher}, \citenamefont {Baumert}, \citenamefont {Richter}, \citenamefont {Kronj{\"a}ger}, \citenamefont {Bongs},\ and\ \citenamefont {Sengstock}}]{becker2008oscillations}%
  \BibitemOpen
  \bibfield  {author} {\bibinfo {author} {\bibfnamefont {C.}~\bibnamefont {Becker}}, \bibinfo {author} {\bibfnamefont {S.}~\bibnamefont {Stellmer}}, \bibinfo {author} {\bibfnamefont {P.}~\bibnamefont {Soltan-Panahi}}, \bibinfo {author} {\bibfnamefont {S.}~\bibnamefont {D{\"o}rscher}}, \bibinfo {author} {\bibfnamefont {M.}~\bibnamefont {Baumert}}, \bibinfo {author} {\bibfnamefont {E.-M.}\ \bibnamefont {Richter}}, \bibinfo {author} {\bibfnamefont {J.}~\bibnamefont {Kronj{\"a}ger}}, \bibinfo {author} {\bibfnamefont {K.}~\bibnamefont {Bongs}},\ and\ \bibinfo {author} {\bibfnamefont {K.}~\bibnamefont {Sengstock}},\ }\href@noop {} {\bibfield  {journal} {\bibinfo  {journal} {Nature Physics}\ }\textbf {\bibinfo {volume} {4}},\ \bibinfo {pages} {496} (\bibinfo {year} {2008})}\BibitemShut {NoStop}%
\bibitem [{\citenamefont {Kuznetsov}\ \emph {et~al.}(1986)\citenamefont {Kuznetsov}, \citenamefont {Rubenchik},\ and\ \citenamefont {Zakharov}}]{kuznetsov1986soliton}%
  \BibitemOpen
  \bibfield  {author} {\bibinfo {author} {\bibfnamefont {E.}~\bibnamefont {Kuznetsov}}, \bibinfo {author} {\bibfnamefont {A.}~\bibnamefont {Rubenchik}},\ and\ \bibinfo {author} {\bibfnamefont {V.~E.}\ \bibnamefont {Zakharov}},\ }\href@noop {} {\bibfield  {journal} {\bibinfo  {journal} {Physics Reports}\ }\textbf {\bibinfo {volume} {142}},\ \bibinfo {pages} {103} (\bibinfo {year} {1986})}\BibitemShut {NoStop}%
\bibitem [{\citenamefont {Pitaevskii}\ and\ \citenamefont {Stringari}(2016)}]{pitaevskii2016bose}%
  \BibitemOpen
  \bibfield  {author} {\bibinfo {author} {\bibfnamefont {L.}~\bibnamefont {Pitaevskii}}\ and\ \bibinfo {author} {\bibfnamefont {S.}~\bibnamefont {Stringari}},\ }\href@noop {} {\emph {\bibinfo {title} {Bose-Einstein condensation and superfluidity}}},\ Vol.\ \bibinfo {volume} {164}\ (\bibinfo  {publisher} {Oxford University Press},\ \bibinfo {year} {2016})\BibitemShut {NoStop}%
\bibitem [{\citenamefont {Schweikhard}\ \emph {et~al.}(2004)\citenamefont {Schweikhard}, \citenamefont {Coddington}, \citenamefont {Engels}, \citenamefont {Tung}, \citenamefont {Cornell},\ and\ \citenamefont {EA}}]{schweikhard2004vortex}%
  \BibitemOpen
  \bibfield  {author} {\bibinfo {author} {\bibfnamefont {V.}~\bibnamefont {Schweikhard}}, \bibinfo {author} {\bibfnamefont {I.}~\bibnamefont {Coddington}}, \bibinfo {author} {\bibfnamefont {P.}~\bibnamefont {Engels}}, \bibinfo {author} {\bibfnamefont {S.}~\bibnamefont {Tung}}, \bibinfo {author} {\bibnamefont {Cornell}},\ and\ \bibinfo {author} {\bibnamefont {EA}},\ }\href@noop {} {\bibfield  {journal} {\bibinfo  {journal} {Physical review letters}\ }\textbf {\bibinfo {volume} {93}},\ \bibinfo {pages} {210403} (\bibinfo {year} {2004})}\BibitemShut {NoStop}%
\bibitem [{\citenamefont {Zhao}(2025)}]{zhao2025spatiotemporal}%
  \BibitemOpen
  \bibfield  {author} {\bibinfo {author} {\bibfnamefont {M.}~\bibnamefont {Zhao}},\ }\href@noop {} {\bibfield  {journal} {\bibinfo  {journal} {Physical Review A}\ }\textbf {\bibinfo {volume} {111}},\ \bibinfo {pages} {033320} (\bibinfo {year} {2025})}\BibitemShut {NoStop}%
\bibitem [{\citenamefont {Zhao}\ \emph {et~al.}(2025)\citenamefont {Zhao}, \citenamefont {Tao},\ and\ \citenamefont {Spielman}}]{zhao2025kolmogorov}%
  \BibitemOpen
  \bibfield  {author} {\bibinfo {author} {\bibfnamefont {M.}~\bibnamefont {Zhao}}, \bibinfo {author} {\bibfnamefont {J.}~\bibnamefont {Tao}},\ and\ \bibinfo {author} {\bibfnamefont {I.}~\bibnamefont {Spielman}},\ }\href@noop {} {\bibfield  {journal} {\bibinfo  {journal} {Physical Review Letters}\ }\textbf {\bibinfo {volume} {134}},\ \bibinfo {pages} {083402} (\bibinfo {year} {2025})}\BibitemShut {NoStop}%
\bibitem [{\citenamefont {Bao}\ \emph {et~al.}(2006)\citenamefont {Bao}, \citenamefont {Chern},\ and\ \citenamefont {Lim}}]{bao2006efficient}%
  \BibitemOpen
  \bibfield  {author} {\bibinfo {author} {\bibfnamefont {W.}~\bibnamefont {Bao}}, \bibinfo {author} {\bibfnamefont {I.-L.}\ \bibnamefont {Chern}},\ and\ \bibinfo {author} {\bibfnamefont {F.~Y.}\ \bibnamefont {Lim}},\ }\href@noop {} {\bibfield  {journal} {\bibinfo  {journal} {Journal of Computational Physics}\ }\textbf {\bibinfo {volume} {219}},\ \bibinfo {pages} {836} (\bibinfo {year} {2006})}\BibitemShut {NoStop}%
\bibitem [{\citenamefont {Chang}\ and\ \citenamefont {Chien}(2007)}]{chang2007adaptive}%
  \BibitemOpen
  \bibfield  {author} {\bibinfo {author} {\bibfnamefont {S.-L.}\ \bibnamefont {Chang}}\ and\ \bibinfo {author} {\bibfnamefont {C.-S.}\ \bibnamefont {Chien}},\ }\href@noop {} {\bibfield  {journal} {\bibinfo  {journal} {Computer Physics Communications}\ }\textbf {\bibinfo {volume} {177}},\ \bibinfo {pages} {707} (\bibinfo {year} {2007})}\BibitemShut {NoStop}%
\bibitem [{\citenamefont {Chen}\ \emph {et~al.}(2011)\citenamefont {Chen}, \citenamefont {Chern},\ and\ \citenamefont {Wang}}]{chen2011exploring}%
  \BibitemOpen
  \bibfield  {author} {\bibinfo {author} {\bibfnamefont {J.-H.}\ \bibnamefont {Chen}}, \bibinfo {author} {\bibfnamefont {I.-L.}\ \bibnamefont {Chern}},\ and\ \bibinfo {author} {\bibfnamefont {W.}~\bibnamefont {Wang}},\ }\href@noop {} {\bibfield  {journal} {\bibinfo  {journal} {Journal of Computational Physics}\ }\textbf {\bibinfo {volume} {230}},\ \bibinfo {pages} {2222} (\bibinfo {year} {2011})}\BibitemShut {NoStop}%
\bibitem [{\citenamefont {Marojevi{\'c}}\ \emph {et~al.}(2013)\citenamefont {Marojevi{\'c}}, \citenamefont {G{\"o}kl{\"u}},\ and\ \citenamefont {L{\"a}mmerzahl}}]{marojevic2013energy}%
  \BibitemOpen
  \bibfield  {author} {\bibinfo {author} {\bibfnamefont {{\v{Z}}.}~\bibnamefont {Marojevi{\'c}}}, \bibinfo {author} {\bibfnamefont {E.}~\bibnamefont {G{\"o}kl{\"u}}},\ and\ \bibinfo {author} {\bibfnamefont {C.}~\bibnamefont {L{\"a}mmerzahl}},\ }\href@noop {} {\bibfield  {journal} {\bibinfo  {journal} {Computer Physics Communications}\ }\textbf {\bibinfo {volume} {184}},\ \bibinfo {pages} {1920} (\bibinfo {year} {2013})}\BibitemShut {NoStop}%
\bibitem [{\citenamefont {Liu}\ \emph {et~al.}(2023)\citenamefont {Liu}, \citenamefont {Xie},\ and\ \citenamefont {Yuan}}]{liu2023constrained}%
  \BibitemOpen
  \bibfield  {author} {\bibinfo {author} {\bibfnamefont {W.}~\bibnamefont {Liu}}, \bibinfo {author} {\bibfnamefont {Z.}~\bibnamefont {Xie}},\ and\ \bibinfo {author} {\bibfnamefont {Y.}~\bibnamefont {Yuan}},\ }\href@noop {} {\bibfield  {journal} {\bibinfo  {journal} {Journal of Computational Physics}\ }\textbf {\bibinfo {volume} {473}},\ \bibinfo {pages} {111719} (\bibinfo {year} {2023})}\BibitemShut {NoStop}%
\bibitem [{\citenamefont {Medvidovi{\'c}}\ and\ \citenamefont {Moreno}(2024)}]{medvidovic2024neural}%
  \BibitemOpen
  \bibfield  {author} {\bibinfo {author} {\bibfnamefont {M.}~\bibnamefont {Medvidovi{\'c}}}\ and\ \bibinfo {author} {\bibfnamefont {J.~R.}\ \bibnamefont {Moreno}},\ }\href@noop {} {\bibfield  {journal} {\bibinfo  {journal} {The European Physical Journal Plus}\ }\textbf {\bibinfo {volume} {139}},\ \bibinfo {pages} {1} (\bibinfo {year} {2024})}\BibitemShut {NoStop}%
\bibitem [{\citenamefont {Lange}\ \emph {et~al.}(2024)\citenamefont {Lange}, \citenamefont {Van~de Walle}, \citenamefont {Abedinnia},\ and\ \citenamefont {Bohrdt}}]{lange2024architectures}%
  \BibitemOpen
  \bibfield  {author} {\bibinfo {author} {\bibfnamefont {H.}~\bibnamefont {Lange}}, \bibinfo {author} {\bibfnamefont {A.}~\bibnamefont {Van~de Walle}}, \bibinfo {author} {\bibfnamefont {A.}~\bibnamefont {Abedinnia}},\ and\ \bibinfo {author} {\bibfnamefont {A.}~\bibnamefont {Bohrdt}},\ }\href@noop {} {\bibfield  {journal} {\bibinfo  {journal} {Quantum Science and Technology}\ } (\bibinfo {year} {2024})}\BibitemShut {NoStop}%
\bibitem [{\citenamefont {Guti{\'e}rrez}\ and\ \citenamefont {Mendl}(2022)}]{gutierrez2022real}%
  \BibitemOpen
  \bibfield  {author} {\bibinfo {author} {\bibfnamefont {I.~L.}\ \bibnamefont {Guti{\'e}rrez}}\ and\ \bibinfo {author} {\bibfnamefont {C.~B.}\ \bibnamefont {Mendl}},\ }\href@noop {} {\bibfield  {journal} {\bibinfo  {journal} {Quantum}\ }\textbf {\bibinfo {volume} {6}},\ \bibinfo {pages} {627} (\bibinfo {year} {2022})}\BibitemShut {NoStop}%
\bibitem [{\citenamefont {Bao}\ \emph {et~al.}(2025)\citenamefont {Bao}, \citenamefont {Chang},\ and\ \citenamefont {Zhao}}]{bao2025computing}%
  \BibitemOpen
  \bibfield  {author} {\bibinfo {author} {\bibfnamefont {W.}~\bibnamefont {Bao}}, \bibinfo {author} {\bibfnamefont {Z.}~\bibnamefont {Chang}},\ and\ \bibinfo {author} {\bibfnamefont {X.}~\bibnamefont {Zhao}},\ }\href@noop {} {\bibfield  {journal} {\bibinfo  {journal} {Journal of Computational Physics}\ }\textbf {\bibinfo {volume} {520}},\ \bibinfo {pages} {113486} (\bibinfo {year} {2025})}\BibitemShut {NoStop}%
\bibitem [{\citenamefont {Poudel}\ \emph {et~al.}(2024)\citenamefont {Poudel}, \citenamefont {Wang},\ and\ \citenamefont {Lee}}]{poudel2024novel}%
  \BibitemOpen
  \bibfield  {author} {\bibinfo {author} {\bibfnamefont {S.}~\bibnamefont {Poudel}}, \bibinfo {author} {\bibfnamefont {X.}~\bibnamefont {Wang}},\ and\ \bibinfo {author} {\bibfnamefont {S.}~\bibnamefont {Lee}},\ }\href@noop {} {\bibfield  {journal} {\bibinfo  {journal} {Engineering Applications of Artificial Intelligence}\ }\textbf {\bibinfo {volume} {133}},\ \bibinfo {pages} {108313} (\bibinfo {year} {2024})}\BibitemShut {NoStop}%
\bibitem [{\citenamefont {Morton}\ and\ \citenamefont {Mayers}(2005)}]{morton2005numerical}%
  \BibitemOpen
  \bibfield  {author} {\bibinfo {author} {\bibfnamefont {K.~W.}\ \bibnamefont {Morton}}\ and\ \bibinfo {author} {\bibfnamefont {D.~F.}\ \bibnamefont {Mayers}},\ }\href@noop {} {\emph {\bibinfo {title} {Numerical solution of partial differential equations: an introduction}}}\ (\bibinfo  {publisher} {Cambridge university press},\ \bibinfo {year} {2005})\BibitemShut {NoStop}%
\bibitem [{\citenamefont {Johnson}(2009)}]{johnson2009numerical}%
  \BibitemOpen
  \bibfield  {author} {\bibinfo {author} {\bibfnamefont {C.}~\bibnamefont {Johnson}},\ }\href@noop {} {\emph {\bibinfo {title} {Numerical solution of partial differential equations by the finite element method}}}\ (\bibinfo  {publisher} {Courier Corporation},\ \bibinfo {year} {2009})\BibitemShut {NoStop}%
\bibitem [{\citenamefont {Galerkin}(1915)}]{galerkin1915sterzhni}%
  \BibitemOpen
  \bibfield  {author} {\bibinfo {author} {\bibfnamefont {B.}~\bibnamefont {Galerkin}},\ }\href@noop {} {\bibfield  {journal} {\bibinfo  {journal} {Vestnik inzhenerov}\ }\textbf {\bibinfo {volume} {1}},\ \bibinfo {pages} {897} (\bibinfo {year} {1915})}\BibitemShut {NoStop}%
\bibitem [{\citenamefont {Bland}\ \emph {et~al.}(2018)\citenamefont {Bland}, \citenamefont {Parker}, \citenamefont {Proukakis},\ and\ \citenamefont {Malomed}}]{bland2018probing}%
  \BibitemOpen
  \bibfield  {author} {\bibinfo {author} {\bibfnamefont {T.}~\bibnamefont {Bland}}, \bibinfo {author} {\bibfnamefont {N.~G.}\ \bibnamefont {Parker}}, \bibinfo {author} {\bibfnamefont {N.~P.}\ \bibnamefont {Proukakis}},\ and\ \bibinfo {author} {\bibfnamefont {B.~A.}\ \bibnamefont {Malomed}},\ }\href@noop {} {\bibfield  {journal} {\bibinfo  {journal} {Journal of Physics B: Atomic, Molecular and Optical Physics}\ }\textbf {\bibinfo {volume} {51}},\ \bibinfo {pages} {205303} (\bibinfo {year} {2018})}\BibitemShut {NoStop}%
\bibitem [{\citenamefont {Podivilov}\ \emph {et~al.}(2019)\citenamefont {Podivilov}, \citenamefont {Kharenko}, \citenamefont {Gonta}, \citenamefont {Krupa}, \citenamefont {Sidelnikov}, \citenamefont {Turitsyn}, \citenamefont {Fedoruk}, \citenamefont {Babin},\ and\ \citenamefont {Wabnitz}}]{podivilov2019hydrodynamic}%
  \BibitemOpen
  \bibfield  {author} {\bibinfo {author} {\bibfnamefont {E.~V.}\ \bibnamefont {Podivilov}}, \bibinfo {author} {\bibfnamefont {D.~S.}\ \bibnamefont {Kharenko}}, \bibinfo {author} {\bibfnamefont {V.}~\bibnamefont {Gonta}}, \bibinfo {author} {\bibfnamefont {K.}~\bibnamefont {Krupa}}, \bibinfo {author} {\bibfnamefont {O.~S.}\ \bibnamefont {Sidelnikov}}, \bibinfo {author} {\bibfnamefont {S.}~\bibnamefont {Turitsyn}}, \bibinfo {author} {\bibfnamefont {M.~P.}\ \bibnamefont {Fedoruk}}, \bibinfo {author} {\bibfnamefont {S.~A.}\ \bibnamefont {Babin}},\ and\ \bibinfo {author} {\bibfnamefont {S.}~\bibnamefont {Wabnitz}},\ }\href@noop {} {\bibfield  {journal} {\bibinfo  {journal} {Physical review letters}\ }\textbf {\bibinfo {volume} {122}},\ \bibinfo {pages} {103902} (\bibinfo {year} {2019})}\BibitemShut {NoStop}%
\bibitem [{\citenamefont {Rumelhart}\ \emph {et~al.}(1986)\citenamefont {Rumelhart}, \citenamefont {Hinton},\ and\ \citenamefont {Williams}}]{rumelhart1986learning}%
  \BibitemOpen
  \bibfield  {author} {\bibinfo {author} {\bibfnamefont {D.~E.}\ \bibnamefont {Rumelhart}}, \bibinfo {author} {\bibfnamefont {G.~E.}\ \bibnamefont {Hinton}},\ and\ \bibinfo {author} {\bibfnamefont {R.~J.}\ \bibnamefont {Williams}},\ }\href@noop {} {\bibfield  {journal} {\bibinfo  {journal} {nature}\ }\textbf {\bibinfo {volume} {323}},\ \bibinfo {pages} {533} (\bibinfo {year} {1986})}\BibitemShut {NoStop}%
\bibitem [{\citenamefont {Bottou}(2010)}]{bottou2010large}%
  \BibitemOpen
  \bibfield  {author} {\bibinfo {author} {\bibfnamefont {L.}~\bibnamefont {Bottou}},\ }in\ \href@noop {} {\emph {\bibinfo {booktitle} {Proceedings of COMPSTAT'2010: 19th International Conference on Computational StatisticsParis France, August 22-27, 2010 Keynote, Invited and Contributed Papers}}}\ (\bibinfo {organization} {Springer},\ \bibinfo {year} {2010})\ pp.\ \bibinfo {pages} {177--186}\BibitemShut {NoStop}%
\bibitem [{\citenamefont {Kingma}\ and\ \citenamefont {Ba}(2014)}]{kingma2014adam}%
  \BibitemOpen
  \bibfield  {author} {\bibinfo {author} {\bibfnamefont {D.~P.}\ \bibnamefont {Kingma}}\ and\ \bibinfo {author} {\bibfnamefont {J.}~\bibnamefont {Ba}},\ }\href@noop {} {\bibfield  {journal} {\bibinfo  {journal} {arXiv preprint arXiv:1412.6980}\ } (\bibinfo {year} {2014})}\BibitemShut {NoStop}%
\bibitem [{\citenamefont {Baydin}\ \emph {et~al.}(2018)\citenamefont {Baydin}, \citenamefont {Pearlmutter}, \citenamefont {Radul},\ and\ \citenamefont {Siskind}}]{baydin2018automatic}%
  \BibitemOpen
  \bibfield  {author} {\bibinfo {author} {\bibfnamefont {A.~G.}\ \bibnamefont {Baydin}}, \bibinfo {author} {\bibfnamefont {B.~A.}\ \bibnamefont {Pearlmutter}}, \bibinfo {author} {\bibfnamefont {A.~A.}\ \bibnamefont {Radul}},\ and\ \bibinfo {author} {\bibfnamefont {J.~M.}\ \bibnamefont {Siskind}},\ }\href@noop {} {\bibfield  {journal} {\bibinfo  {journal} {Journal of machine learning research}\ }\textbf {\bibinfo {volume} {18}},\ \bibinfo {pages} {1} (\bibinfo {year} {2018})}\BibitemShut {NoStop}%
\bibitem [{\citenamefont {Stokes}\ \emph {et~al.}(2020)\citenamefont {Stokes}, \citenamefont {Izaac}, \citenamefont {Killoran},\ and\ \citenamefont {Carleo}}]{stokes2020quantum}%
  \BibitemOpen
  \bibfield  {author} {\bibinfo {author} {\bibfnamefont {J.}~\bibnamefont {Stokes}}, \bibinfo {author} {\bibfnamefont {J.}~\bibnamefont {Izaac}}, \bibinfo {author} {\bibfnamefont {N.}~\bibnamefont {Killoran}},\ and\ \bibinfo {author} {\bibfnamefont {G.}~\bibnamefont {Carleo}},\ }\href@noop {} {\bibfield  {journal} {\bibinfo  {journal} {Quantum}\ }\textbf {\bibinfo {volume} {4}},\ \bibinfo {pages} {269} (\bibinfo {year} {2020})}\BibitemShut {NoStop}%
\bibitem [{\citenamefont {Amari}(2016)}]{amari2016information}%
  \BibitemOpen
  \bibfield  {author} {\bibinfo {author} {\bibfnamefont {S.-i.}\ \bibnamefont {Amari}},\ }\href@noop {} {\emph {\bibinfo {title} {Information geometry and its applications}}},\ Vol.\ \bibinfo {volume} {194}\ (\bibinfo  {publisher} {Springer},\ \bibinfo {year} {2016})\BibitemShut {NoStop}%
\bibitem [{\citenamefont {Martens}\ and\ \citenamefont {Grosse}(2015)}]{martens2015optimizing}%
  \BibitemOpen
  \bibfield  {author} {\bibinfo {author} {\bibfnamefont {J.}~\bibnamefont {Martens}}\ and\ \bibinfo {author} {\bibfnamefont {R.}~\bibnamefont {Grosse}},\ }in\ \href@noop {} {\emph {\bibinfo {booktitle} {International conference on machine learning}}}\ (\bibinfo {organization} {PMLR},\ \bibinfo {year} {2015})\ pp.\ \bibinfo {pages} {2408--2417}\BibitemShut {NoStop}%
\bibitem [{\citenamefont {Paszke}\ \emph {et~al.}(2017)\citenamefont {Paszke}, \citenamefont {Gross}, \citenamefont {Chintala}, \citenamefont {Chanan}, \citenamefont {Yang}, \citenamefont {DeVito}, \citenamefont {Lin}, \citenamefont {Desmaison}, \citenamefont {Antiga},\ and\ \citenamefont {Lerer}}]{paszke2017automatic}%
  \BibitemOpen
  \bibfield  {author} {\bibinfo {author} {\bibfnamefont {A.}~\bibnamefont {Paszke}}, \bibinfo {author} {\bibfnamefont {S.}~\bibnamefont {Gross}}, \bibinfo {author} {\bibfnamefont {S.}~\bibnamefont {Chintala}}, \bibinfo {author} {\bibfnamefont {G.}~\bibnamefont {Chanan}}, \bibinfo {author} {\bibfnamefont {E.}~\bibnamefont {Yang}}, \bibinfo {author} {\bibfnamefont {Z.}~\bibnamefont {DeVito}}, \bibinfo {author} {\bibfnamefont {Z.}~\bibnamefont {Lin}}, \bibinfo {author} {\bibfnamefont {A.}~\bibnamefont {Desmaison}}, \bibinfo {author} {\bibfnamefont {L.}~\bibnamefont {Antiga}},\ and\ \bibinfo {author} {\bibfnamefont {A.}~\bibnamefont {Lerer}},\ }\href@noop {} {\  (\bibinfo {year} {2017})}\BibitemShut {NoStop}%
\bibitem [{\citenamefont {Hendrycks}\ and\ \citenamefont {Gimpel}(2016)}]{hendrycks2016gaussian}%
  \BibitemOpen
  \bibfield  {author} {\bibinfo {author} {\bibfnamefont {D.}~\bibnamefont {Hendrycks}}\ and\ \bibinfo {author} {\bibfnamefont {K.}~\bibnamefont {Gimpel}},\ }\href@noop {} {\bibfield  {journal} {\bibinfo  {journal} {arXiv preprint arXiv:1606.08415}\ } (\bibinfo {year} {2016})}\BibitemShut {NoStop}%
\bibitem [{\citenamefont {Zwiebach}(2022)}]{zwiebach2022mastering}%
  \BibitemOpen
  \bibfield  {author} {\bibinfo {author} {\bibfnamefont {B.}~\bibnamefont {Zwiebach}},\ }\href@noop {} {\emph {\bibinfo {title} {Mastering quantum mechanics: essentials, theory, and applications}}}\ (\bibinfo  {publisher} {MIT Press},\ \bibinfo {year} {2022})\BibitemShut {NoStop}%
\bibitem [{\citenamefont {Cranmer}(2023)}]{cranmer2023interpretable}%
  \BibitemOpen
  \bibfield  {author} {\bibinfo {author} {\bibfnamefont {M.}~\bibnamefont {Cranmer}},\ }\href@noop {} {\bibfield  {journal} {\bibinfo  {journal} {arXiv preprint arXiv:2305.01582}\ } (\bibinfo {year} {2023})}\BibitemShut {NoStop}%
\bibitem [{\citenamefont {Makke}\ and\ \citenamefont {Chawla}(2024)}]{makke2024interpretable}%
  \BibitemOpen
  \bibfield  {author} {\bibinfo {author} {\bibfnamefont {N.}~\bibnamefont {Makke}}\ and\ \bibinfo {author} {\bibfnamefont {S.}~\bibnamefont {Chawla}},\ }\href@noop {} {\bibfield  {journal} {\bibinfo  {journal} {Artificial Intelligence Review}\ }\textbf {\bibinfo {volume} {57}},\ \bibinfo {pages} {2} (\bibinfo {year} {2024})}\BibitemShut {NoStop}%
\bibitem [{\citenamefont {Cranmer}\ \emph {et~al.}(2020)\citenamefont {Cranmer}, \citenamefont {Sanchez~Gonzalez}, \citenamefont {Battaglia}, \citenamefont {Xu}, \citenamefont {Cranmer}, \citenamefont {Spergel},\ and\ \citenamefont {Ho}}]{cranmer2020discovering}%
  \BibitemOpen
  \bibfield  {author} {\bibinfo {author} {\bibfnamefont {M.}~\bibnamefont {Cranmer}}, \bibinfo {author} {\bibfnamefont {A.}~\bibnamefont {Sanchez~Gonzalez}}, \bibinfo {author} {\bibfnamefont {P.}~\bibnamefont {Battaglia}}, \bibinfo {author} {\bibfnamefont {R.}~\bibnamefont {Xu}}, \bibinfo {author} {\bibfnamefont {K.}~\bibnamefont {Cranmer}}, \bibinfo {author} {\bibfnamefont {D.}~\bibnamefont {Spergel}},\ and\ \bibinfo {author} {\bibfnamefont {S.}~\bibnamefont {Ho}},\ }\href@noop {} {\bibfield  {journal} {\bibinfo  {journal} {Advances in neural information processing systems}\ }\textbf {\bibinfo {volume} {33}},\ \bibinfo {pages} {17429} (\bibinfo {year} {2020})}\BibitemShut {NoStop}%
\bibitem [{\citenamefont {Liu}\ \emph {et~al.}(2024{\natexlab{a}})\citenamefont {Liu}, \citenamefont {Wang}, \citenamefont {Vaidya}, \citenamefont {Ruehle}, \citenamefont {Halverson}, \citenamefont {Solja{\v{c}}i{\'c}}, \citenamefont {Hou},\ and\ \citenamefont {Tegmark}}]{liu2024kan}%
  \BibitemOpen
  \bibfield  {author} {\bibinfo {author} {\bibfnamefont {Z.}~\bibnamefont {Liu}}, \bibinfo {author} {\bibfnamefont {Y.}~\bibnamefont {Wang}}, \bibinfo {author} {\bibfnamefont {S.}~\bibnamefont {Vaidya}}, \bibinfo {author} {\bibfnamefont {F.}~\bibnamefont {Ruehle}}, \bibinfo {author} {\bibfnamefont {J.}~\bibnamefont {Halverson}}, \bibinfo {author} {\bibfnamefont {M.}~\bibnamefont {Solja{\v{c}}i{\'c}}}, \bibinfo {author} {\bibfnamefont {T.~Y.}\ \bibnamefont {Hou}},\ and\ \bibinfo {author} {\bibfnamefont {M.}~\bibnamefont {Tegmark}},\ }\href@noop {} {\bibfield  {journal} {\bibinfo  {journal} {arXiv preprint arXiv:2404.19756}\ } (\bibinfo {year} {2024}{\natexlab{a}})}\BibitemShut {NoStop}%
\bibitem [{\citenamefont {Liu}\ \emph {et~al.}(2024{\natexlab{b}})\citenamefont {Liu}, \citenamefont {Ma}, \citenamefont {Wang}, \citenamefont {Matusik},\ and\ \citenamefont {Tegmark}}]{liu2024kan2}%
  \BibitemOpen
  \bibfield  {author} {\bibinfo {author} {\bibfnamefont {Z.}~\bibnamefont {Liu}}, \bibinfo {author} {\bibfnamefont {P.}~\bibnamefont {Ma}}, \bibinfo {author} {\bibfnamefont {Y.}~\bibnamefont {Wang}}, \bibinfo {author} {\bibfnamefont {W.}~\bibnamefont {Matusik}},\ and\ \bibinfo {author} {\bibfnamefont {M.}~\bibnamefont {Tegmark}},\ }\href@noop {} {\bibfield  {journal} {\bibinfo  {journal} {arXiv preprint arXiv:2408.10205}\ } (\bibinfo {year} {2024}{\natexlab{b}})}\BibitemShut {NoStop}%
\bibitem [{\citenamefont {Kolmogorov}(1961)}]{kolmogorov1961representation}%
  \BibitemOpen
  \bibfield  {author} {\bibinfo {author} {\bibfnamefont {A.~N.}\ \bibnamefont {Kolmogorov}},\ }\href@noop {} {\emph {\bibinfo {title} {On the representation of continuous functions of several variables by superpositions of continuous functions of a smaller number of variables}}}\ (\bibinfo  {publisher} {American Mathematical Society},\ \bibinfo {year} {1961})\BibitemShut {NoStop}%
\bibitem [{\citenamefont {De~Boor}\ and\ \citenamefont {De~Boor}(1978)}]{de1978practical}%
  \BibitemOpen
  \bibfield  {author} {\bibinfo {author} {\bibfnamefont {C.}~\bibnamefont {De~Boor}}\ and\ \bibinfo {author} {\bibfnamefont {C.}~\bibnamefont {De~Boor}},\ }\href@noop {} {\emph {\bibinfo {title} {A practical guide to splines}}},\ Vol.~\bibinfo {volume} {27}\ (\bibinfo  {publisher} {springer New York},\ \bibinfo {year} {1978})\BibitemShut {NoStop}%
\bibitem [{\citenamefont {Takahashi}\ \emph {et~al.}(2015)\citenamefont {Takahashi}, \citenamefont {Kobayashi},\ and\ \citenamefont {Nitta}}]{takahashi2015nambu}%
  \BibitemOpen
  \bibfield  {author} {\bibinfo {author} {\bibfnamefont {D.~A.}\ \bibnamefont {Takahashi}}, \bibinfo {author} {\bibfnamefont {M.}~\bibnamefont {Kobayashi}},\ and\ \bibinfo {author} {\bibfnamefont {M.}~\bibnamefont {Nitta}},\ }\href@noop {} {\bibfield  {journal} {\bibinfo  {journal} {Physical Review B}\ }\textbf {\bibinfo {volume} {91}},\ \bibinfo {pages} {184501} (\bibinfo {year} {2015})}\BibitemShut {NoStop}%
\bibitem [{\citenamefont {Indekeu}\ \emph {et~al.}(2015)\citenamefont {Indekeu}, \citenamefont {Lin}, \citenamefont {Van~Thu}, \citenamefont {Van~Schaeybroeck},\ and\ \citenamefont {Phat}}]{indekeu2015static}%
  \BibitemOpen
  \bibfield  {author} {\bibinfo {author} {\bibfnamefont {J.~O.}\ \bibnamefont {Indekeu}}, \bibinfo {author} {\bibfnamefont {C.-Y.}\ \bibnamefont {Lin}}, \bibinfo {author} {\bibfnamefont {N.}~\bibnamefont {Van~Thu}}, \bibinfo {author} {\bibfnamefont {B.}~\bibnamefont {Van~Schaeybroeck}},\ and\ \bibinfo {author} {\bibfnamefont {T.~H.}\ \bibnamefont {Phat}},\ }\href@noop {} {\bibfield  {journal} {\bibinfo  {journal} {Physical Review A}\ }\textbf {\bibinfo {volume} {91}},\ \bibinfo {pages} {033615} (\bibinfo {year} {2015})}\BibitemShut {NoStop}%
\bibitem [{\citenamefont {Kolmogorov}()}]{kolmogorov1995turbulence}%
  \BibitemOpen
  \bibfield  {author} {\bibinfo {author} {\bibfnamefont {A.}~\bibnamefont {Kolmogorov}},\ }\href@noop {} {\emph {\bibinfo {title} {Turbulence: the legacy of AN Kolmogorov}}}\BibitemShut {NoStop}%
\bibitem [{\citenamefont {Benzi}\ \emph {et~al.}(1993)\citenamefont {Benzi}, \citenamefont {Ciliberto}, \citenamefont {Tripiccione}, \citenamefont {Baudet}, \citenamefont {Massaioli},\ and\ \citenamefont {Succi}}]{benzi1993extended}%
  \BibitemOpen
  \bibfield  {author} {\bibinfo {author} {\bibfnamefont {R.}~\bibnamefont {Benzi}}, \bibinfo {author} {\bibfnamefont {S.}~\bibnamefont {Ciliberto}}, \bibinfo {author} {\bibfnamefont {R.}~\bibnamefont {Tripiccione}}, \bibinfo {author} {\bibfnamefont {C.}~\bibnamefont {Baudet}}, \bibinfo {author} {\bibfnamefont {F.}~\bibnamefont {Massaioli}},\ and\ \bibinfo {author} {\bibfnamefont {S.}~\bibnamefont {Succi}},\ }\href@noop {} {\bibfield  {journal} {\bibinfo  {journal} {Physical Review E}\ }\textbf {\bibinfo {volume} {48}},\ \bibinfo {pages} {R29} (\bibinfo {year} {1993})}\BibitemShut {NoStop}%
\bibitem [{\citenamefont {Strogatz}(2018)}]{strogatz2018nonlinear}%
  \BibitemOpen
  \bibfield  {author} {\bibinfo {author} {\bibfnamefont {S.~H.}\ \bibnamefont {Strogatz}},\ }\href@noop {} {\emph {\bibinfo {title} {Nonlinear dynamics and chaos: with applications to physics, biology, chemistry, and engineering}}}\ (\bibinfo  {publisher} {CRC press},\ \bibinfo {year} {2018})\BibitemShut {NoStop}%
\bibitem [{\citenamefont {B{\v{r}}ezinov{\'a}}\ \emph {et~al.}(2012)\citenamefont {B{\v{r}}ezinov{\'a}}, \citenamefont {Lode}, \citenamefont {Streltsov}, \citenamefont {Alon}, \citenamefont {Cederbaum},\ and\ \citenamefont {Burgd{\"o}rfer}}]{bvrezinova2012wave}%
  \BibitemOpen
  \bibfield  {author} {\bibinfo {author} {\bibfnamefont {I.}~\bibnamefont {B{\v{r}}ezinov{\'a}}}, \bibinfo {author} {\bibfnamefont {A.~U.}\ \bibnamefont {Lode}}, \bibinfo {author} {\bibfnamefont {A.~I.}\ \bibnamefont {Streltsov}}, \bibinfo {author} {\bibfnamefont {O.~E.}\ \bibnamefont {Alon}}, \bibinfo {author} {\bibfnamefont {L.~S.}\ \bibnamefont {Cederbaum}},\ and\ \bibinfo {author} {\bibfnamefont {J.}~\bibnamefont {Burgd{\"o}rfer}},\ }\href@noop {} {\bibfield  {journal} {\bibinfo  {journal} {Physical Review A}\ }\textbf {\bibinfo {volume} {86}},\ \bibinfo {pages} {013630} (\bibinfo {year} {2012})}\BibitemShut {NoStop}%
\bibitem [{\citenamefont {Ehrenfest}(1927)}]{ehrenfest1927bemerkung}%
  \BibitemOpen
  \bibfield  {author} {\bibinfo {author} {\bibfnamefont {P.}~\bibnamefont {Ehrenfest}},\ }\href@noop {} {\bibfield  {journal} {\bibinfo  {journal} {Zeitschrift f{\"u}r physik}\ }\textbf {\bibinfo {volume} {45}},\ \bibinfo {pages} {455} (\bibinfo {year} {1927})}\BibitemShut {NoStop}%
\bibitem [{\citenamefont {Berman}\ and\ \citenamefont {Zaslavsky}(1978)}]{berman1978condition}%
  \BibitemOpen
  \bibfield  {author} {\bibinfo {author} {\bibfnamefont {G.~P.}\ \bibnamefont {Berman}}\ and\ \bibinfo {author} {\bibfnamefont {G.~M.}\ \bibnamefont {Zaslavsky}},\ }\href@noop {} {\bibfield  {journal} {\bibinfo  {journal} {Physica A: Statistical Mechanics and its Applications}\ }\textbf {\bibinfo {volume} {91}},\ \bibinfo {pages} {450} (\bibinfo {year} {1978})}\BibitemShut {NoStop}%
\bibitem [{\citenamefont {Han}\ and\ \citenamefont {Wu}(2016)}]{han2016ehrenfest}%
  \BibitemOpen
  \bibfield  {author} {\bibinfo {author} {\bibfnamefont {X.}~\bibnamefont {Han}}\ and\ \bibinfo {author} {\bibfnamefont {B.}~\bibnamefont {Wu}},\ }\href@noop {} {\bibfield  {journal} {\bibinfo  {journal} {Physical Review A}\ }\textbf {\bibinfo {volume} {93}},\ \bibinfo {pages} {023621} (\bibinfo {year} {2016})}\BibitemShut {NoStop}%
\bibitem [{\citenamefont {Rammensee}\ \emph {et~al.}(2018)\citenamefont {Rammensee}, \citenamefont {Urbina},\ and\ \citenamefont {Richter}}]{rammensee2018many}%
  \BibitemOpen
  \bibfield  {author} {\bibinfo {author} {\bibfnamefont {J.}~\bibnamefont {Rammensee}}, \bibinfo {author} {\bibfnamefont {J.~D.}\ \bibnamefont {Urbina}},\ and\ \bibinfo {author} {\bibfnamefont {K.}~\bibnamefont {Richter}},\ }\href@noop {} {\bibfield  {journal} {\bibinfo  {journal} {Physical Review Letters}\ }\textbf {\bibinfo {volume} {121}},\ \bibinfo {pages} {124101} (\bibinfo {year} {2018})}\BibitemShut {NoStop}%
\bibitem [{\citenamefont {Wanzenb{\"o}ck}\ \emph {et~al.}(2021)\citenamefont {Wanzenb{\"o}ck}, \citenamefont {Donsa}, \citenamefont {Hofst{\"a}tter}, \citenamefont {Koch}, \citenamefont {Schlagheck},\ and\ \citenamefont {B{\v{r}}ezinov{\'a}}}]{wanzenbock2021chaos}%
  \BibitemOpen
  \bibfield  {author} {\bibinfo {author} {\bibfnamefont {R.}~\bibnamefont {Wanzenb{\"o}ck}}, \bibinfo {author} {\bibfnamefont {S.}~\bibnamefont {Donsa}}, \bibinfo {author} {\bibfnamefont {H.}~\bibnamefont {Hofst{\"a}tter}}, \bibinfo {author} {\bibfnamefont {O.}~\bibnamefont {Koch}}, \bibinfo {author} {\bibfnamefont {P.}~\bibnamefont {Schlagheck}},\ and\ \bibinfo {author} {\bibfnamefont {I.}~\bibnamefont {B{\v{r}}ezinov{\'a}}},\ }\href@noop {} {\bibfield  {journal} {\bibinfo  {journal} {Physical Review A}\ }\textbf {\bibinfo {volume} {103}},\ \bibinfo {pages} {023336} (\bibinfo {year} {2021})}\BibitemShut {NoStop}%
\bibitem [{\citenamefont {Kohn}(1961)}]{kohn1961cyclotron}%
  \BibitemOpen
  \bibfield  {author} {\bibinfo {author} {\bibfnamefont {W.}~\bibnamefont {Kohn}},\ }\href@noop {} {\bibfield  {journal} {\bibinfo  {journal} {Physical Review}\ }\textbf {\bibinfo {volume} {123}},\ \bibinfo {pages} {1242} (\bibinfo {year} {1961})}\BibitemShut {NoStop}%
\bibitem [{\citenamefont {Brey}\ \emph {et~al.}(1989)\citenamefont {Brey}, \citenamefont {Johnson},\ and\ \citenamefont {Halperin}}]{brey1989optical}%
  \BibitemOpen
  \bibfield  {author} {\bibinfo {author} {\bibfnamefont {L.}~\bibnamefont {Brey}}, \bibinfo {author} {\bibfnamefont {N.}~\bibnamefont {Johnson}},\ and\ \bibinfo {author} {\bibfnamefont {B.}~\bibnamefont {Halperin}},\ }\href@noop {} {\bibfield  {journal} {\bibinfo  {journal} {Physical Review B}\ }\textbf {\bibinfo {volume} {40}},\ \bibinfo {pages} {10647} (\bibinfo {year} {1989})}\BibitemShut {NoStop}%
\bibitem [{\citenamefont {Dobson}(1994)}]{dobson1994harmonic}%
  \BibitemOpen
  \bibfield  {author} {\bibinfo {author} {\bibfnamefont {J.~F.}\ \bibnamefont {Dobson}},\ }\href@noop {} {\bibfield  {journal} {\bibinfo  {journal} {Physical Review Letters}\ }\textbf {\bibinfo {volume} {73}},\ \bibinfo {pages} {2244} (\bibinfo {year} {1994})}\BibitemShut {NoStop}%
\bibitem [{\citenamefont {Reynolds}(1895)}]{reynolds1895iv}%
  \BibitemOpen
  \bibfield  {author} {\bibinfo {author} {\bibfnamefont {O.}~\bibnamefont {Reynolds}},\ }\href@noop {} {\bibfield  {journal} {\bibinfo  {journal} {Philosophical transactions of the royal society of london.(a.)}\ ,\ \bibinfo {pages} {123}} (\bibinfo {year} {1895})}\BibitemShut {NoStop}%
\bibitem [{\citenamefont {Kolmogorov}(1941)}]{kolmogorov1941degeneration}%
  \BibitemOpen
  \bibfield  {author} {\bibinfo {author} {\bibfnamefont {A.~N.}\ \bibnamefont {Kolmogorov}},\ }in\ \href@noop {} {\emph {\bibinfo {booktitle} {Dokl. Akad. Nauk SSSR}}},\ Vol.~\bibinfo {volume} {31}\ (\bibinfo {year} {1941})\ pp.\ \bibinfo {pages} {538--540}\BibitemShut {NoStop}%
\bibitem [{\citenamefont {Kolmogorov}(1991{\natexlab{a}})}]{kolmogorov1991local}%
  \BibitemOpen
  \bibfield  {author} {\bibinfo {author} {\bibfnamefont {A.~N.}\ \bibnamefont {Kolmogorov}},\ }\href@noop {} {\bibfield  {journal} {\bibinfo  {journal} {Proceedings of the Royal Society of London. Series A: Mathematical and Physical Sciences}\ }\textbf {\bibinfo {volume} {434}},\ \bibinfo {pages} {9} (\bibinfo {year} {1991}{\natexlab{a}})}\BibitemShut {NoStop}%
\bibitem [{\citenamefont {Kolmogorov}(1991{\natexlab{b}})}]{kolmogorov1991dissipation}%
  \BibitemOpen
  \bibfield  {author} {\bibinfo {author} {\bibfnamefont {A.~N.}\ \bibnamefont {Kolmogorov}},\ }\href@noop {} {\bibfield  {journal} {\bibinfo  {journal} {Proceedings of the Royal Society of London. Series A: Mathematical and Physical Sciences}\ }\textbf {\bibinfo {volume} {434}},\ \bibinfo {pages} {15} (\bibinfo {year} {1991}{\natexlab{b}})}\BibitemShut {NoStop}%
\bibitem [{\citenamefont {Schmid}(2010)}]{schmid2010dynamic}%
  \BibitemOpen
  \bibfield  {author} {\bibinfo {author} {\bibfnamefont {P.~J.}\ \bibnamefont {Schmid}},\ }\href@noop {} {\bibfield  {journal} {\bibinfo  {journal} {Journal of fluid mechanics}\ }\textbf {\bibinfo {volume} {656}},\ \bibinfo {pages} {5} (\bibinfo {year} {2010})}\BibitemShut {NoStop}%
\bibitem [{\citenamefont {Schmid}(2022)}]{schmid2022dynamic}%
  \BibitemOpen
  \bibfield  {author} {\bibinfo {author} {\bibfnamefont {P.~J.}\ \bibnamefont {Schmid}},\ }\href@noop {} {\bibfield  {journal} {\bibinfo  {journal} {Annual Review of Fluid Mechanics}\ }\textbf {\bibinfo {volume} {54}},\ \bibinfo {pages} {225} (\bibinfo {year} {2022})}\BibitemShut {NoStop}%
\bibitem [{\citenamefont {Thomas}\ \emph {et~al.}(2021)\citenamefont {Thomas}, \citenamefont {Davis},\ and\ \citenamefont {Kheruntsyan}}]{thomas2021thermalization}%
  \BibitemOpen
  \bibfield  {author} {\bibinfo {author} {\bibfnamefont {K.~F.}\ \bibnamefont {Thomas}}, \bibinfo {author} {\bibfnamefont {M.~J.}\ \bibnamefont {Davis}},\ and\ \bibinfo {author} {\bibfnamefont {K.~V.}\ \bibnamefont {Kheruntsyan}},\ }\href@noop {} {\bibfield  {journal} {\bibinfo  {journal} {Physical Review A}\ }\textbf {\bibinfo {volume} {103}},\ \bibinfo {pages} {023315} (\bibinfo {year} {2021})}\BibitemShut {NoStop}%
\bibitem [{\citenamefont {Orlova}\ \emph {et~al.}(2023)\citenamefont {Orlova}, \citenamefont {Ustimenko}, \citenamefont {Jiang}, \citenamefont {Lu},\ and\ \citenamefont {Willett}}]{orlova2023deep}%
  \BibitemOpen
  \bibfield  {author} {\bibinfo {author} {\bibfnamefont {E.}~\bibnamefont {Orlova}}, \bibinfo {author} {\bibfnamefont {A.}~\bibnamefont {Ustimenko}}, \bibinfo {author} {\bibfnamefont {R.}~\bibnamefont {Jiang}}, \bibinfo {author} {\bibfnamefont {P.~Y.}\ \bibnamefont {Lu}},\ and\ \bibinfo {author} {\bibfnamefont {R.}~\bibnamefont {Willett}},\ }\href@noop {} {\bibfield  {journal} {\bibinfo  {journal} {arXiv preprint arXiv:2305.19685}\ } (\bibinfo {year} {2023})}\BibitemShut {NoStop}%
\bibitem [{\citenamefont {Geng}\ \emph {et~al.}(2024)\citenamefont {Geng}, \citenamefont {Tao}, \citenamefont {Zhao}, \citenamefont {Mukherjee}, \citenamefont {Eckel}, \citenamefont {Campbell},\ and\ \citenamefont {Spielman}}]{geng2024rayleigh}%
  \BibitemOpen
  \bibfield  {author} {\bibinfo {author} {\bibfnamefont {Y.}~\bibnamefont {Geng}}, \bibinfo {author} {\bibfnamefont {J.}~\bibnamefont {Tao}}, \bibinfo {author} {\bibfnamefont {M.}~\bibnamefont {Zhao}}, \bibinfo {author} {\bibfnamefont {S.}~\bibnamefont {Mukherjee}}, \bibinfo {author} {\bibfnamefont {S.}~\bibnamefont {Eckel}}, \bibinfo {author} {\bibfnamefont {G.~K.}\ \bibnamefont {Campbell}},\ and\ \bibinfo {author} {\bibfnamefont {I.~B.}\ \bibnamefont {Spielman}},\ }\href@noop {} {\bibfield  {journal} {\bibinfo  {journal} {arXiv preprint arXiv:2411.19807}\ } (\bibinfo {year} {2024})}\BibitemShut {NoStop}%
\end{thebibliography}
%

\end{document}